\documentclass[12pt]{iopart}
\usepackage{iopams}
\usepackage{graphicx}

\begin{document}

\title[Free-fall of photons in a planar optical cavity]{Free-fall of photons in a planar optical cavity}

\author{Maxime Richard}

\address{Univ. Grenoble Alpes, CNRS, Grenoble INP, Institut N\'eel, Fr-38000 Grenoble, France}
\vspace{10pt}
\begin{indented}
\item[]
\end{indented}

\begin{abstract}
We consider a planar optical cavity with its main axis oriented perpendicular to the earth's local gravity field. We examine the motion of a wavepacket of light trapped within the cavity spacer under the action of gravity. Within a relativistic framework, we find that the wavepacket initially at rest in terms of in-plane group velocity is accelerated by $g$ at early time in agreement the non-relativistic limit of free-fall. We show that this phenomenon is a manifestation of gravitational redshift, applied on photons at rest within the cavity.
\end{abstract}

\section{INTRODUCTION}

The effect of gravity on propagating electromagnetic waves leads to several fascinating phenomena such as gravitational lensing \cite{walsh_1979}, redshift \cite{pound_1960} and time dilation \cite{shapiro_1999}. These effects are often associated with light propagating freely over macroscopic distances; and yet, they can also show up in photonic devices designed for this purpose such as interferometers. Let us mention for instance the Sagnac interferometer routinely used nowadays in inertial navigation \cite{post_1967}, and the celebrated giant Michelson interferometers used in the LIGO-Virgo collaborations, that detected recently several gravitational wavetrains from black-holes and neutron stars mergers \cite{abbott_1,abbott_2,abbott_3}.

Another class of photonic device is less frequently considered and yet increasingly relevant in this context: that of optical resonators (i.e. Fabry-Perot optical cavities). These devices are capable of storing a wavepacket of electromagnetic field within a small volume for a certain time. In state-of-the-art cavities, this storage time has become strikingly large: In the near-visible domain, photons have been stored for several microseconds in whispering-gallery modes in solid-state cavities of large diameters \cite{lee_2012,hofer_2010,farnesi_2014,goryachev_2014}, up to a record lifetime of $50\,\mu$s \cite{grudinin_2006}. In the microwave domain, photons can be stored nowadays in a confocal superconducting cavity for as long as $0.13\,$s \cite{gleyzes_2007} which, once unfolded, correspond to a free space travel as large as $39000\,$km. Such space and time scales suggest that the field stored in these cavity might be influenced by earth's gravity in a measurable way.

\begin{figure}[t]
\centering
\includegraphics[width=0.7\textwidth]{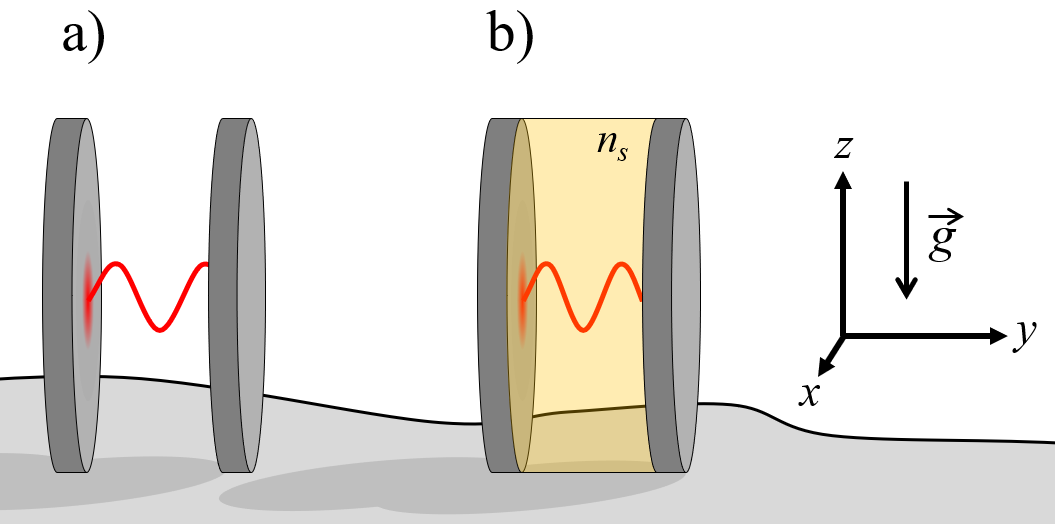}
\caption{Pictorial representation of the system considered in this work. The two mirrors of a planar Fabry-Perot cavity are shown standing vertically, i.e. parallel to the earth's gravity field $g$, assuming an empty spacer in a) as considered in sections \ref{section1} and \ref{section2}, and a spacer filled by a solid-state material of refractive index $n_s$ in b), as considered in sections \ref{section3} and \ref{section4}. The axes convention and the direction of the gravity field are shown on the side. The size of the mirrors and the thickness of the spacer are not at scale (see section \ref{section4} for a realistic experimental implementation).}
\label{fig1}
\end{figure}

With this idea in mind, a subclass of optical cavities is of particular conceptual interest: the large size ones that exhibit effective translational invariance along one or two dimension (2D). This is for instance the case of semiconductor planar optical cavities \cite{sun_2017}, or of optical-fiber-like cylindrical waveguides.  We will focus in the next three sections ont the two-dimensional (2D) case, without lack of generality. In a planar optical cavity, the kinetic properties of cavity photons maps accurately that of 2D massive particles freely moving in this 2D space \cite{chiao_1999}; Indeed, the textbook derivation of the cavity mode dispersion relation can be written as $\hbar\omega(\mathbf{p}_\parallel)=\sqrt{|\mathbf{p}_\parallel|^2c^2 + m_\parallel^2c^4}$, that involves a well-defined rest-energy and kinetic-energy terms associated to this in-plane motion, where $m_\parallel=\hbar\omega_0/c^2$ is the effective photon mass that, unlike in free space, correctly describes its rest energy, $\omega_0/2\pi$ is the cavity resonance at $k_\parallel=0$, $\mathbf{k}_\parallel=\mathbf{p}_\parallel/\hbar$ is the in-plane wavenumber, and $\mathbf{p}_\parallel$ the in-plane momentum. Note that there are other longitudinal modes present in the cavity that potentially break up this description. However when the modes are well split from each others (i.e. large Finesse cavity), and for low wavenumber $k_\parallel$, photons in a given longitudinal mode number $j$ cannot jump into the neighbouring modes $j\pm 1$ without breaking momentum and energy conservation.

In this work, in the first three sections, we examine theoretically the motion of a wavepacket of light living within the spacer of such a planar cavity, when at least one of the two free dimensions is aligned with the gravity field vector $\mathbf{g}$. In agreement with the equivalence principle, and upon taking care of applying it only along the translationally invariant directions, we show that the wavepacket of light undergoes free fall within the cavity spacer. Since photons are effectively stopped and put at rest with respect to the laboratory frame by the cavity, this free fall motion has a non-relativistic regime at low group velocity that matches our everyday life experience of the Newtonian free fall. We show that this behaviour can be understood as a manifestation of gravitational redshift in a cavity.

\section{Gravitational free-fall in an empty cavity}
\label{section1}


The setup which is considered is that depicted in Fig.\ref{fig1}.a: a planar optical cavity with infinite size mirrors (i.e. much larger than the in-plane size of the trapped wavepacket of light. See section \ref{section4} for details) is lying at rest in the laboratory frame at the surface of the earth, with its main axis ($y$) parallel with the horizon, and one of the in-plane axis ($z$) oriented parallel to $\mathbf{g}$. The cavity embeds a long lived wavepacket of light which is initially at rest in terms of its in-plane group velocity $v_g^z(t=0)=\hbar k_\parallel/m_\parallel=0$ in the laboratory frame.

Before going further, two properties that stem from the translational invariance of the cavity need to be underlined as they will prove crucial: (1) An observer situated outside the cavity can measure the energy and momentum of the photons inside the cavity, using the photons leaking throughout the imperfectly reflecting mirrors. Indeed, both the energy and the in-plane momentum of the light is conserved in this inside/outside weak coupling mechanism as a result of the cavity translational invariance \cite{houdre_1994}. (2) In the equivalence principle that can be formulated for this system, the motion  along ($z$) or ($x$) of the two mirrors forming the cavity is irrelevant as it leaves the problem invariant. As a result, the wavepacket of light within the cavity can be considered at will either in free fall or at rest by an observer, independently from the mirrors in-plane motion. Note that in this idealized model, we disregard the light induced electronic excitations within the metallic material constituting the mirrors, otherwise statement (2) would not hold. We will see in the section \ref{section3} how having a light-excited (polarizable) material in the cavity indeed breaks down the equivalence principle.

With these precisions in mind, we can formulate the equivalence principle between the light within the empty cavity and the observer as follows: (i) an accelerated observer (say at $g$, along $z$) with respect to the intracavity wavepacket, who measures its energy and in-plane momentum will obtain the same result as (ii) an observer at rest performing the same measurement on an accelerated intracavity wavepacket (say at $g$, along $z$). (ii) is the experimental situation, and (i) shows us immediately that the observer is going to see the a build up of a vertical in-plane momentum of the wavepacket over time, at a rate to be determined, but fixed by $\mathbf{g}$.

In order to get a quantitative derivation of this gravity induced dynamics, we determine the intracavity dispersion relation in a general relativity framework. Since we restrict ourselves to a small intracavity volume $\mathcal{V}$, and elevation above the surface of the earth, as compared to the Schwarzschild metric curvature, we can use the Newtonian limit of the metric with the line element \cite{plebanski_2006}:
\begin{equation}
ds_\mathcal{V}^2=-\left[1-r_s/(r_t+z)\right]c^2dt^2+\left[1-r_s/(r_t+z)\right]^{-1}dz^2+dx^2+dy^2.
\end{equation}
where $r_s=2GM_t/c^2=2gr_t^2/c^2 \ll r_t$ is the earth Schwarzschild radius, $r_t$ is the earth radius, $M_t$ is the earth's mass, $G$ is the gravitational constant, and $g=9.81\,$m.s$^{-2}$ is the gravitational acceleration at the surface of the earth. The vertical axis $z$ is the vertical axis, and $z=0$ is set at the surface of the earth ($z\ll r_t$). We will see in the last section that the in-plane size of $\mathcal{V}$ that fixes the mirror radius, is essentially fixed by the intracavity in-plane expansion of the wavepacket resulting from diffraction. $\mathcal{V}$ is determined such that the fraction of the field reaching the mirror edges is negligible at any time during the cavity storage lifetime.

In order to derive the dispersion relation, we adopt a technique based on the Lagragian action principle developed in \cite{kulsrud_1992}, which has the advantage of properly accounting for a possible polarizability of the medium, as will be done in section \ref{section3}. In the present section, we consider an empty cavity. Using the aforementioned elements for the metric tensor, the transverse field dispersion relation reads
\begin{equation}
f_T=-g^{00}\omega^2+c^2(k_z^2 g^{zz}+ [k^j_y]^2+k_x^2)=0,
\end{equation}
where
\begin{equation}
g^{00}(z)=g^{zz}(z)^{-1}=1-\frac{r_s}{r_t+z}\equiv F(z), \label{Fdez}
\end{equation}
and $\mathbf{k}_\parallel=(k_x,k_z)$. Since the metric is assumed flat along the cavity axis $y$, i.e. $g^{yy}=1$, the cavity thickness $L_c$ does not depend on the observer height $z$. The constructive interference along the cavity main axis thus leads to $k^j_y=j\pi/L_c$ where $j$ is fixed the longitudinal mode number. Note that we have assumed a scalar field for the sake of simplicity, and thus neglect in this work a potentially rich, but much more complicated physics pertaining to the polarization degree of freedom. This approximation is well justified here, as the electromagnetic field oscillation plane tilt angle $ck_\parallel/\omega_0$  with respect to the mirror plane remains very small at all time.

 The dispersion relation can thus be rewritten as
\begin{equation}
f_T=-g^{00}(z)\omega^2+c^2(k_z^2 g^{zz}(z)+m_\parallel^2 c^2/\hbar^2)=0, \label{ft}
\end{equation}
where $k_x=0$ is assumed for simplicity and without loss of generality, and $m_\parallel$ is defined like in the previous section. This expression is the dispersion relation of light within the cavity, accounting for the gravitational correction. The group velocity of the wavepacket can be derived from eq.(\ref{ft}) \cite{kulsrud_1992} as
\begin{equation}
v_g^z=\dot{z}=\frac{\partial f_T}{\partial k_z}\left[\frac{\partial f_T}{\partial \omega}\right]^{-1}. \label{vggrav}
\end{equation}
A relation on the time derivative of the wavevector $k_z$ can also be derived, resulting from the fact that the wavepacket is subject to gravitational acceleration:
\begin{equation}
\dot{k_z}=-\frac{\partial f_T}{\partial z}\left[\frac{\partial f_T}{\partial \omega}\right]^{-1}. \label{kpgrav}
\end{equation}
Equations (\ref{ft}), (\ref{vggrav}) and (\ref{kpgrav}) provide a complete description of the wavepacket dynamics inside the cavity plane. We can for instance calculate the acceleration of the wavepacket within the cavity. To do so, we take the time derivative of eq.(\ref{vggrav}), and express it as a function of $v_g^z$, and $k_z$. We obtain a lengthy expression that assumes a simple nonrelativistic approximation, namely
\begin{equation}
a_g^z=\ddot{z}=\frac{1}{2}\frac{c^2}{F^3}\frac{\partial F}{\partial z} + \mathcal{O}\left(\frac{c^2k_z^2}{\omega^2} \right). \label{av}
\end{equation}
where according to eq.(\ref{vggrav}), $\epsilon=\dot{z}/c \simeq ck_z/\omega \ll 1$ and $F\simeq 1$ are the usual nonrelativistic (Newtonian) approximation. In this limit, this expression simplifies into
\begin{equation}
a_g^z=-g,
\end{equation}
up to the second order in $\epsilon$. This is the usual Newtonian limit describing the free fall of a finite mass object.

We thus see that a wavepacket of light confined in a cavity plane with an initially vanishing in-plane group velocity, will fall towards the ground with the same acceleration $g$ as a any solid object, without qualitative nor quantitative distinction. This derivation also show that our 2D-restricted equivalence principle is well preserved, even in this limiting situation where the particle 2D effective inertial mass emerges from the confinement in the cavity plane. While we did not verify it, we expect this equivalence principle to hold in the relativistic regime as well.

\section{Relation with the gravitational redshift}
\label{section2}

The phenomenon of gravitational redshift results from the fact that a free falling observer measuring the energy of photons in a light beam at two separate points in space, should obtain the same result. Therefore if the observer is not joining these two points by free falling, a spectral shift will be found, which is reminiscent of a doppler shift.

This phenomenon bears a significant resemblance with our situation, except that since we are dealing with a wavepacket of light at rest (in the cavity) instead of propagating in free space at $c$, the two measurements considered above take place at two different points in time instead of space. We can check further this interpretation by comparing the change rate of photons momentum in both cases. In our case, expanding Eq.(\ref{kpgrav}) in the non-relativistic limit yields
\begin{equation}
\dot{k_z}=-\frac{g\omega_0}{c^2} \label{kz_dot},
\end{equation}
which is the rate of in-plane momentum increase (with a negative sign as it points 'downward') per time units due to the gravitational field.

The gravitational redshift reads $\delta\omega_{\rm rs}/\omega_0=g\delta z/c^2$ \cite{plebanski_2006}, as obtained in the limit of weak gravitational field and low height $\delta z$. Since we consider events separated in time instead of space, we can replace $\delta z$ by $\delta t=\delta z/c$, such that $\delta\omega_{\rm rs}/\omega_0=g\delta t/c$, and $\delta\omega_{\rm rs}=c\delta k_{z,{\rm rs}}$. We thus find that
\begin{equation}
\dot{k}_{z,{\rm rs}}=-\frac{g\omega_0}{c^2} \label{kz_rs_dot},
\end{equation}
which is identical to that derived in the intracavity free fall motion that we discussed.

\section{Gravitational free-fall in a solid-state cavity}
\label{section3}
For practical purposes, we wish to examine how this free fall motion is modified when the cavity spacer is filled with a solid-state polarizable medium of refractive index $n_s>1$ as depicted in Fig.\ref{fig1}.b, which is the case of most state-of-the-art optical cavities in the near-visible domain. To describe this situation, the perturbation of the electrons rest position in the medium is added to the action to be minimized \cite{kulsrud_1992}. In our specific geometry, this method yields the following dispersion relation :
\begin{equation}
f^s_T=(1-g^{00}(z)-n^2_s)\omega^2+c^2(k_z^2 g^{zz}(z)+m_\parallel^2 c^2/\hbar)=0.
\label{fts}
\end{equation}
Note that due to the prefactor in front of $\omega$, the inertial mass is modified from its vacuum expression into $m_\parallel^s=m_\parallel\times n_s$, while the speed of light changes into $c_s=c/n_s$. We now carry out the same derivation as in section \ref{section1} to determine the dynamics of the wavepacket in this polarizable medium. Upon applying the Newtonian approximation, we end up with the following gravity-induced acceleration of the wavepacket
\begin{equation}
a_g^z=\ddot{z}=\frac{1}{2}\frac{c^2}{F(1-n_s^2-F)^2}\frac{\partial F}{\partial z} + \mathcal{O}\left(\frac{c^2k_z^2}{\omega^2} \right),
\label{av2}
\end{equation}
that simplifies into
\begin{equation}
a_g^z=-g/n_s^4,
\end{equation}
up to the second order in $\epsilon$ and $F(1-n_s^2-F)^2 \simeq n_s^4$. Thus, upon filling the cavity spacer with a polarizable material, our wavepacket of light still undergoes a Newtonian free fall but with a slowed down acceleration $g_s=g/n_s^4$. This feature can be understood by the fact that the electronic motion excited by the intracavity light and generating the polarization field are fixed in the cavity frame and thus are not in free fall, unlike the electromagnetic field. The earlier thus backacts on the field as a kind of drag force as pointed out already in the context of light propagation in transparent moving media \cite{leonhardt_1999}, which thus breaks up the equivalence principle.

Note that when $n_s$ depends on the frequency (in a so-called dispersive medium), the notion of effective mass, or in-plane rest mass, associated with the in-plane wavepackets motion, also breaks down as this feature has the effect of altering the dispersion relation. A paradigmatic examples of this situation is the strong coupling regime in planar semiconductor microcavities \cite{carusotto_2013}, in which $n_s$ modifies the dispersion relation so much that it opens a spectral gap in the vicinity of the bound electron-hole pair transition.

\section{Experimental signature}
\label{section4}

\begin{figure}[t]
\centering
\includegraphics[width=0.5\textwidth]{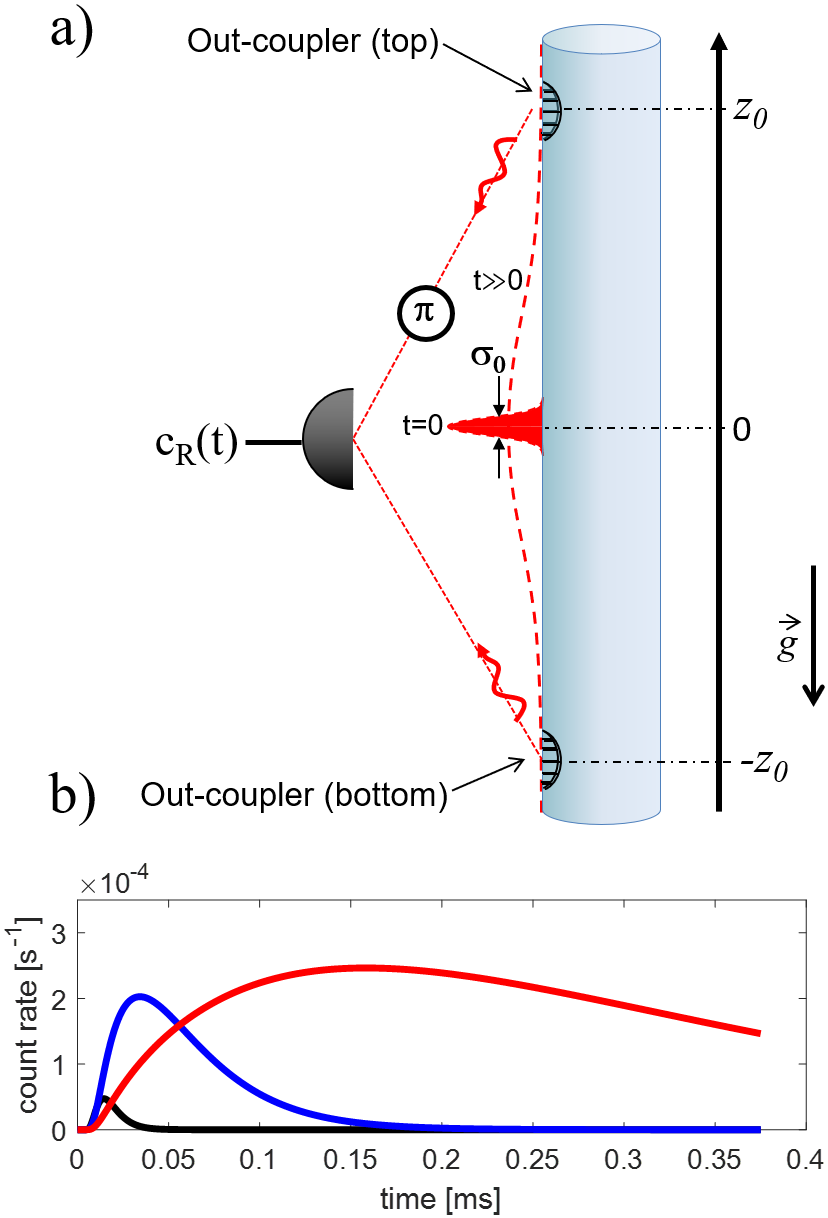}
\caption{a) Experimental setup for the measurement of Newtonian free fall in a cylindrical optical cavity. The WGM cavity has a diameter in the millimeter range in order to reach giant quality factor \cite{grudinin_2006}. b) calculated count rate $\dot{c}_R(t)$ in as measured on the photodiode, following the injection of a short light pulse at $t=0$, assuming $Q=0.2\times 10^{10}$ (black), $Q=1\times 10^{10}$ (blue), $Q=6\times 10^{10}$ (red). For this calculation we used a wavelength of $\lambda=780\,$nm, an initial waist of the injected WGM $\sigma_0=5\,$mm; the out-coupling height is $z_0=10\,$cm, refractive index $n_s=1.43$ (CaF$_2$), and outcoupler surface factor $s_{oc}=5\%$.}
\label{fig2}
\end{figure}


In this section, we examine whether this free fall is actually observable in state-of-the-art optical cavities. To carry out this measurement, we can exploit the fact that, unlike classical solid bodies, we have access to the phase $\phi(z,t)$ of the wavepacket, of which $k_z$ is the gradient along $z$. Its dynamic, is given by eq.(\ref{kz_dot}) in a vacuum cavity. Using eq.(\ref{kpgrav}) and eq.(\ref{fts}) to account for the refractive index $n_s$ of the cavity medium,
\begin{equation}
\dot{k_z}=-\frac{g\omega_0}{n_s^2c^2}=-\frac{g_s\omega_0}{c_s^2} \label{kz_dot_s},
\end{equation}
which is thus slower than in vacuum.

In both cases, the gravitational field builds-up a phase gradient over time that can be detected into an interferometric measurement. Assuming a quality factor $Q=1\times 10^{10}$, the phase gradient achieved within the intracavity lifetime is of the order of $10^{-6}\,$rad.m$^{-1}$. This is a very small phase shift, that might prove challenging to detect, but not impossible. A phase resolution $<10^{-4}\,$rad is typically achievable in well-controlled but standard interferometric setup \cite{hosseini_2016}.

Let us examine a possible implementation of such a measurement: owing to their record performance in the visible domain, we choose to consider a solid-state cylindrical cavity as depicted in Fig.\ref{fig2}.a. Based on results from several groups \cite{lee_2012,hofer_2010,grudinin_2006}, similar structures made up of ultra pure materials like CaF$_2$ or MgF$_2$ support whispering gallery mode (WGM) with a quality factor as large as $Q=6.1\times10^{10}$ \cite{grudinin_2006}. Note that the results we have derived so far can be transposed directly from the planar into the cylindrical cavity. In cylindrical cavity, the light is confined into whispering gallery modes in the transverse direction, and free to propagate along the main axis (with a scalar momentum $k_a$), which is thus oriented vertically along $z$.

The experiment would be carried out in the following way: at $t=0$ a spatially Gaussian pulse of light (of initial waist $\sigma_0$ along $z$), is injected in the middle of the cylindrical cavity with an initial momentum $\hbar k_a=0$, in the chosen whispering gallery mode by evanescent coupling. The free evolving wavepacket is then going to be subject to two different dynamics. The first one has nothing to do with gravity, and results from the laws of diffraction: the mode waist $\sigma(t)$ is going to expand in both directions of the WGM cavity as $\sigma^2(t)=\sigma^2_0(1+[\lambda ct/(\pi\sigma_0^2n_s^2)^2])$ \cite{bornwolf}. We take advantage of this effect to couple the light out of the WGM only after a singificant delay fixed by this expansion, using grating-like diffracting elements placed at both ends of the waveguide (in $z_0$ and $-z_0$). Note that in absence of gravity this free expansion does not introduce any phase difference between ($z$) and ($-z$) direction of the expansion.

The second effect is the one we are looking for, it introduces a slowly varying phase difference between the two beams outcoupled at $z_0$ and $-z_0$. The two beams are sent to interfere on a state-of-the-art single photon detector with high quantum efficiency, such as an avalanche photodiode. A phase retardation is inserted into one of the arms of the interferometer such that at $t=0$ the phase difference between both paths is exactly $\pi$, and the interference is perfectly destructive, except for the effect of gravity.

The photon counts on the photodiode, can be simulated within a simple rate equation model where
\begin{equation}
\dot{N}=-[\gamma_c+\gamma_{oc}s_{oc}\alpha(t)]N(t),
\label{signal}
\end{equation}
is the loss rate of the photons outside the WGM cavity; $\gamma_c=\omega_0/Q$ is the nominal cavity loss rate, $\gamma_{oc}$ is the outcoupling rate throughout the diffractive elements. $\alpha(t)$ is the overlap integral between the gaussian mode and the diffractive elements, and $s_{oc}$ is a surface factor characterizing these elements size. The count rate on the photodiode then reads
\begin{equation}
\dot{c}_R=\gamma_{oc}s_{oc}\alpha(t)[1+\cos{\Delta\phi(t)}]N(t),
\label{signal}
\end{equation}
where $\Delta\phi(z_0,t)=2z_0\omega_0g_sn_s^2t/c_s^2$ is the gravity induced time-dependent phase shift between points $z_0$ and $-z_0$ given by eq.(\ref{kz_dot_s}). We solved this model numerically using the realistic parameters given in the caption of Fig.\ref{fig2}. Examples of calculated $\dot{c}_R(t)$ are shown in Fig.\ref{fig2}.b for $Q=\{0.2,1,6.0\}\times 10^{10}$. This plot shows how gravity induces a slow phase shift in the interferometer than eventually lets some light hit the detector. In an experiment in which the aim is to demonstrate the free fall phenomenon, we do not need this time resolution, and just need to find a total photon counts $n_T$ that exceed the experimental noise. It can be calculated as $n_T=n_{p}\int{\rm d}t \dot{c}_R(t)$, where $n_p=2.7\times 10^{11}$ is the number of pulses contributing during a 1h integration time. The results are summarized in the following table:
\begin{center}
\begin{tabular}{|c||c|c|c|}
\hline
$Q(\times 10^{10})$&0.2&1.0&6.0\\
\hline
$n_T$&200&3750&19220\\
\hline
$S_n$&14&61&138\\
\hline
\end{tabular}
\end{center}
where $S_n=\sqrt{n_T}$ is the signal-to-noise ratio in a shot-noise-limited measurement. These number are highly encouraging even though it should be kept in mind that in realistic experiments, other source of noises like e.g. the acquisition electronics can spoil $S_n$ significantly. Note that a continuous wave version of this experiment would give a similar result. The major challenge in this experiment is the fabrication of a WGM cavity with negligible fluctuations of the diameter along ($z$), that could otherwise cause the free fall to slow, stop, or even bounce back.

This proposed approach is of course not unique, superconducting cavities holds a record in terms of storage time, that could be exploited using observable that does not depend on the wavelength of light like in an interferometric measurement, like tracking the intensity maximum of the wavepacket over time. in another approach, we could also give up the need for translational invariance of the cavity. Indeed, while we have not considered the case in this work, we expect that gravity couples transverse modes in 3D confined cavity that feature a discrete photonic density of state along ($z$), like microdisks \cite{lee_2012}, microtorus \cite{hofer_2010} or microspheres \cite{farnesi_2014}, or any equivalent system in the microwave domain \cite{goryachev_2014}. In this approach, we expect a slow transfer from one vertical transverse mode to the next as a signature of the free fall.


\section{Conclusion}
In summary, we have shown that when a planar or cylindrical optical cavity is oriented adequately in a gravitational field, an analogous of the gravitational redshift phenomenon takes place. While the 'acceleration' of the wavepacket of light - better expressed in terms of increase rate of momentum - is identical both in free space and in such a cavity, the second case exhibits every features of a nonrelativistic free fall: The wavepacket of light being initially at rest within the cavity, sees its group velocity increase as $gt$ like any solid object of finite mass. In a last section, we propose and analyze an experimental strategy to assess whether this weak effect can be measured in a state-of-the-art optical cavitiy. To do so, we consider a large diameter solid-state cylindrical cavity, and a detection method based on interferometry. Using realistic experimental parameters taken from the literature, we simulate this setup and find that the intracavity free fall motion is in principle observable in a shot-noise limited measurement.

\section*{Acknowledgments}
Discussions with A.Baas, M. Holzmann, O. Arcizet, D. Gerace, I. Carusotto, T.Volz and C. Schneider are warmly acknowledged. This work has been supported by the Australian Research Council eQus network, and by the French Agence National de la Recherche, contract nb ANR-16-CE30-0021 (QFL).


\section*{Bibliography}
\begin{thebibliography}{20}

%
%
\bibitem{walsh_1979} D. Walsh, R. F. Carswell, R. J. Weymann, ``0957 + 561 A, B: twin quasistellar objects or gravitational lens ? '' Nature \textbf{279} 381 (1979)
%
\bibitem{pound_1960} R. V. Pound and G. A. Rebka ``Apparent weight of photons'' Phys. Rev. Lett. \textbf{4} 337 (1960)

\bibitem{shapiro_1999} Irwin I. Shapiro ``A century of relativity'' Rev. Mod. Phys. \textbf{71} (1999) S41

\bibitem{post_1967} E. J. Post, ``Sagnac Effect'', Rev. Mod. Phys. \textbf{39} (1967)

\bibitem{abbott_1} B. P. Abbott et al. (LIGO Scientific Collaboration and Virgo Collaboration), ``Observation of Gravitational Waves from a Binary Black Hole Merger'', Phys. Rev. Lett. \textbf{116}, 061102 (2016)

\bibitem{abbott_2} B. P. Abbott et al. (LIGO Scientific and Virgo Collaboration) ``GW170104: Observation of a 50-Solar-Mass Binary Black Hole Coalescence at Redshift 0.2'' Phys. Rev. Lett. \textbf{118}, 221101 (2017);

\bibitem{abbott_3} B. P. Abbott et al. (LIGO Scientific Collaboration and Virgo Collaboration), ``GW170817: Observation of Gravitational Waves from a Binary Neutron Star Inspiral'' Phys. Rev. Lett. \textbf{119}, 161101 (2017)

\bibitem{lee_2012} H. Lee, T. Chen, J. Li, Ki Y. Yang, S. Jeon, O. Painter and K. J. Vahala ``Chemically etched ultrahigh-Q wedge-resonator on a silicon chip'', Nature Photonics \textbf{6} 369 (2012)

\bibitem{hofer_2010} J. Hofer, A. Schliesser, and T. J. Kippenberg ``Cavity optomechanics with ultrahigh-Q crystalline microresonators'' Phys. Rev. A \textbf{82} 031804(R) (2010)
\bibitem{farnesi_2014} D. Farnesi, A. Barucci, G. C. Righini, S. Berneschi, S. Soria, and G. Nunzi Conti ``Optical Frequency Conversion in Silica-Whispering-Gallery-Mode Microspherical Resonators'' Phys. Rev. Lett. \textbf{112} 093901 (2014)

\bibitem{goryachev_2014} M. Goryachev, W. G. Farr, D. L. Creedon, Y. Fan, M. Kostylev, and M. E. Tobar ``High-Cooperativity Cavity QED with Magnons at Microwave Frequencies''
Phys. Rev. Applied \textbf{2} 054002 (2014).

\bibitem{grudinin_2006} I. S. Grudinin, V. S. Ilchenko and L. Maleki ``Ultrahigh optical Q factors of crystalline resonators in the linear regime'', Phys. Rev. A \textbf{74} 063806 (2006)

\bibitem{gleyzes_2007} S. Gleyzes, S. Kuhr, C. Guerlin, J. Bernu, S. Del\'eglise, U. B. Hoff, M Brune, J.-M. Raimond and S. Haroche, ``Quantum jumps of light recording the birth and death of a photon in a cavity'', Nature \textbf{446} 297 (2007)

\bibitem{sun_2017} Y. Sun, P. Wen, Y. Yoon, G. Liu, M. Steger, L. N. Pfeiffer, K. West, D. W. Snoke, and K. A. Nelson, ``Bose-Einstein Condensation of Long-Lifetime Polaritons in Thermal Equilibrium'' Phys. Rev. Lett. \textbf{118}, 016602 (2017)

\bibitem{chiao_1999} Raymond Y. Chiao and Jack Boyce, ``Bogoliubov dispersion relation and the possibility of superfluidity for weakly interacting photons in a two-dimensional photon fluid'' Phys. Rev. A \textbf{60} 4114 (1999)

%
\bibitem{houdre_1994} R. Houdr\'e, C. Weisbuch, R. P. Stanley, U. Oesterle, P. Pellandini, and M. Ilegems ``Measurement of Cavity-Polariton Dispersion Curve from Angle-Resolved Photoluminescence Experiments'' Phys. Rev. Lett. 2043 \textbf{73} (1994)

\bibitem{plebanski_2006} Jerzy Plebanski and Andrzej Krasinski, ``An Introduction to General Relativity and Cosmology'' edited by Cambridge University Press (2006)

%



\bibitem{kulsrud_1992}
R. Kulsrud and A. Loeb Phys. Rev. D \textbf{45} 525 (1992)

\bibitem{leonhardt_1999}
U. Leonhardt and P. Piwnicki ``Optics of nonuniformly moving media'' Phys. Rev. A \textbf{60} 4301 (1999)
%
\bibitem{carusotto_2013} I. Carusotto and C. Ciuti ``Quantum fluids of light'' Rev. Mod. Phys. \textbf{85} 299 (2013)

\bibitem{hosseini_2016}
P. Hosseini, R. Zhou, Y.-H. Kim, C. Peres, A. Diaspro, C. Kuang, Z. Yaqoob, And P. T. C. So,``Pushing phase and amplitude sensitivity limits in interferometric microscopy'' Optics Letters\textbf{41}, 1656 (2016)

\bibitem{bornwolf}
Max Born, Emil Wolf, ``Principles of Optics: Electromagnetic Theory of Propagation, Interference and Diffraction of Light'', Edited by Elsevier (2013)


%
%
%


%
%
%

\end{thebibliography}
\end{document}